\documentclass[12pt,preprint]{aastex}

\newcommand{\rosat}{{\sl ROSAT\/}}
\newcommand{\asca}{{\sl ASCA\/}}

\newcommand{\xte}{{\sl RXTE\/}}

\newcommand{\xmm}{{\sl XMM-Newton\/}}

\newcommand{\suzaku}{{\sl Suzaku\/}}

\newcommand{\nh}{N$_{\rm H}$}
\newcommand{\cps}{ct\,s$^{-1}$}
\newcommand{\eps}{ergs\,s$^{-1}$}
\newcommand{\epcs}{ergs\,cm$^{-2}$s$^{-1}$}

\newcommand{\vsco}{V893~Sco}

\shorttitle{Partial X-ray Eclipse in V893 Sco}
\shortauthors{Mukai et al.}

\begin{document}


\title{Suzaku Observations of the Dwarf Nova V893 Scorpii:
       the Discovery of a Partial X-ray Eclipse}

 	
\author{K. Mukai}
\affil{CRESST and X-ray Astrophysics Laboratory, NASA/GSFC, Greenbelt,
       MD 20771; and Department of Physics, University of Maryland,
       Baltimore County, 1000 Hilltop Circle, Baltimore, MD 21250.}
\email{Koji.Mukai@nasa.gov}
\author{E. Zietsman}
\affil{Department of Astronomy, University of Cape Town, Private Bag X3,
       Rondebosch 7701, South Africa.}
%\email{ewald.zietsman@gmail.com}

\and
\author{M. Still}
\affil{NASA Ames Research Center, Moffett Field, CA 94035.}
%\email{mds@mssl.ucl.ac.uk}


\altaffiltext{1}{Present address: }


\begin{abstract}
\vsco\ is an eclipsing dwarf nova that had attracted little
attention from X-ray astronomers until it was proposed as the
identification of an \xte\ all-sky slew survey (XSS) source.  Here
we report on the pointed X-ray observations of this
object using \suzaku.  \vsco\ was in quiescence at the time, as
indicated by the coordinated optical photometry we obtained at
the South African Astronomical Observatory.   Our \suzaku\ data
show \vsco\ to be X-ray bright, with a highly absorbed spectrum.
Most importantly, we have discovered a partial X-ray eclipse in \vsco. 
This is the first time that a partial eclipse is seen in X-ray
light curves of a dwarf nova.  Our preliminary simulations demonstrate
that the partial X-ray eclipse can be in principle reproduced if
the white dwarf in \vsco\ is partially eclipsed.
Higher quality observations of this object have the potential to
place significant constraints on the latitudinal extent of the X-ray
emission region and thereby discriminating between an equatorial
boundary layer and a spherical corona.
The partial X-ray eclipse therefore makes \vsco\ a key object in
understanding the physics of accretion in quiescent dwarf nova.

\end{abstract}




\section{Introduction}

Cataclysmic variables (CVs), in which a white dwarf primary accretes
from a Roche-lobe filling, late-type secondary (see \citealt{tome} for
a comprehensive review), are an excellent laboratory for the physics of
accretion.  In the subclass of dwarf novae,  accretion proceeds via a
disk that switches between the low (quiescence) and high (outburst)
luminosity states.  The visible light of a quiescent dwarf nova is
usually dominated by the bright spot, where the accretion flow from
the secondary hits the outer edge of the disk, as well as the photosphere
of the white dwarf.  In outburst, the disk becomes the dominant source
of visible light.  In contrast, X-ray observations of dwarf novae probe
the accretion flow in the immediate vicinity of the white dwarf, such as
an optically thin boundary layer \citep{PR1985a}, since the accretion
disk in a dwarf nova is too cool to emit X-rays.  Although the disk
instability model is a highly successful framework for explaining the
dwarf nova outbursts, there are details that defy the prediction of
the basic version of the model \citep{L2001}.  In particular, the
observed X-ray luminosities of quiescent dwarf nova (often of order
10$^{31}$ \eps; \citealt{Bea2005}) imply an accretion rate onto the
white dwarf that is much higher than predicted.   Proposed modifications
of the disk instability model include the coronal siphon flow \citep{MM1994},
which might lead to accretion over a much wider area of the white
dwarf surface than through a boundary layer, and a weakly magnetic
white dwarf \citep{LP1992}.  In the latter model, the magnetic field
is too weak to control the accretion flow during outburst, but strong
enough to do so during quiescence.

\vsco\ is an eclipsing \citep{BSG2000} dwarf nova with an orbital period
of 1.8 hr \citep{T1999}.  According to \cite{KMU2002}, \vsco\ has a
quiescent magnitude of $\sim$14.5, and has an outburst every $\sim$30 days
during which it reaches magnitude $\sim$12.5.   Extensive photometry
by Warner and collaborators \citep{WWP2003,PWW2006} have revealed
quasi-periodic oscillations and dwarf nova oscillations in \vsco,
but a strictly periodic signal was never found.  Therefore, existing optical
data point strongly towards a dwarf nova classification and argue against
an intermediate polar (IP, or DQ Her type systems; \citealt{P1994})
classification.  IPs are a subset of magnetic CVs
in which the primary's magnetic field disrupts the inner accretion disk,
channeling the flow to the magnetic polar region(s); the spin period of
the magnetic white dwarf is a strict clock that characterize IPs.
Thus, \vsco\ joins a growing number of eclipsing dwarf
novae below the period gap.

Citing the variable shape of the eclipses and the relatively small eclipse
amplitude (often less than 0.75 mag), \cite{BSG2000} argue for a grazing
eclipse of the bright spot and the disk, but not of the white dwarf.  On the
other hand, \cite{Mea2000} argue that the white dwarf is eclipsed, based
on the fact that the spectroscopic conjunction of the disk (the red-to-blue
crossing of the emission line radial velocities) occurs at mid-eclipse.
From the radial velocity curves, they
also infer a white dwarf mass of $\sim$0.5--0.6 M$_\odot$ and a mass ratio
of $\sim$0.2--0.3.  However, \cite{Mea2001} prefer a higher mass
(0.89 M$_\odot$) white dwarf and an inclination angle of 72.5 degrees based
on their analysis of the emission line radial velocities, although this
value depends in part on the assumed mass-radius relationship for the
secondary.

Despite the current lack of a consensus regarding the system
parameters, the eclipsing nature of \vsco\ makes it an important target for
detailed studies.  Moreover, it is a nearby system with
parallax-estimated distance of 155$^{+58}_{-34}$ pc \citep{T2003}.

Although the \rosat\ detection was already noted by \cite{Kea1998},
\vsco\ did not draw the attention of X-ray astronomers until
it was listed among the \xte\ all-sky slew survey (XSS) sources
\citep{Rea2004}, along with three other dwarf novae (SS~Aur,
V426~Oph, and SU~UMa).  The estimated luminosities of these 4 systems
are all just under 10$^{32}$ \eps\ in the 2--10 keV band, placing them
near the upper end for non-magnetic CVs ($10^{30}$--$3 \times 10^{32}$
in 0.1--100 keV for the \asca\ sample; \citealt{Bea2005}).
However, the XSS is based on the data taken during slews of the non-imaging
\xte\ PCA detector.  For relatively faint sources such as these
dwarf novae, the positional errors are considerable, and misidentification
is a possibility.  Hence it is important to confirm the proposed
identification of XSS dwarf novae.  This is particularly true for \vsco,
which had never been the subject of a pointed X-ray observation above 2 keV.
Although \vsco\ is securely detected as a \rosat\ all-sky survey (RASS)
source, the ratio of XSS (2.38 c/s in the 3--8 keV band) to RASS (0.35)
count rates is high, compared to, e.g., SS~Aur which has the same RASS
count rate but was detected at 0.75 c/s in the XSS 3--8 keV band.
This leaves open the possibility that \vsco\ is only a partial contributor
to the XSS flux.  For this reason and also because of the potential return
in studying the X-ray properties of this eclipsing dwarf nova, we performed
a pointed observation using \suzaku, along with contemporaneous optical
photometry.

\section{Observations and Data Reduction}

We observed \vsco\ with \suzaku\ \citep{Mea2007} between 2006 Aug 26
09:40 and Aug 27  01:20 UT (sequence number 401041010).   There are
two types of co-aligned instruments on-board \suzaku, which are:
4 units of the CCD-based X-ray imaging spectrometer (XIS), each behind
its X-ray telescope (XRT), and the non-imaging hard X-ray detector (HXD).
As the source is relatively faint for a non-imaging instrument and the
exposure time short, we did not analyze the HXD data, a decision that
we justify at the end of \S 3.1.

We started from data processed using the V2.0.6.13 pipeline.  We then
updated the energy scale calibration using the 2008 Feb 1 release of the
calibration database.  Initially we applied the standard screening
criteria: attitude control system in nominal mode, pointing
within 1.5 arcmin of the mean direction, XIS data rate medium or high,
the satellite outside the South Atlantic Anomaly (SAA), at least 436 s
after the last SAA passage, elevation above Earth limb $>5^\circ$,
elevation about the bright Earth limb $>20^\circ$.  We then experimented
with relaxing some of the standard criteria to maximize the phase coverage
of the observation.  We found that relaxing the post-SAA passage limit
from 436 s to 180 s, and the bright Earth limb limit from 20$^\circ$
to 15$^\circ$ did not noticeably increase the noise level of our data.
With this modification, the total XIS exposure time was 19,174 s,
representing a 670 s increase compared to the standard screening.
The background level during the screened interval was normal for
\suzaku, and varied within $\pm$50\% of the mean value primarily as
a function of the geomagnetic coordinates of the spacecraft.

We found a bright source centered 13.5 arcsec from the precisely known
optical position of \vsco\ in the XIS image.  Since the current
reconstruction of \suzaku\ attitude results in a 90\% error circle
of 19 arcsec for bright sources \citep{Uea2008}, we identify the
\suzaku\ source with \vsco.  Given the statistical quality of the
image, the apparent offset is caused predominantly by the attitude
reconstruction errors.

We used a 3.5 arcmin radius extraction region for the source and
7.5 and 4 arcmin radius annular extraction region for the background,
both centered on the location of \vsco\ as measured on the XIS
image.  For spectroscopy, we summed the data and the responses of
the three XIS units with frontside illuminated (FI) CCD chips,
because they have nearly identical responses.   The mean net
rate of \vsco\ was 0.85 \cps\ per FI unit, while the estimated background
rate was 0.04 \cps\ in the source region.   The data from XIS1 (with the
backside illuminated, or BI, CCD chip) have a higher background rate
averaging 0.16 \cps, although the net source rate was also higher at 0.98 \cps.
We fit the XIS1 spectrum simultaneously with the sum of FI spectra, after
grouping channels so they have a minimum of 25 counts, thus making
the $\chi^2$ statistic an acceptable approximation.
For photometry, we added background subtracted light curves from all 4
XIS units over the 0.4 keV (FI)/0.3 keV (BI) to 10 keV bandpass using
8 s bins, preserving the original time resolution of the XIS data.
We also extracted light curves below 2 keV, in the 2--4 keV band,
and above 4 keV to investigate possible energy dependence.

We also carried out optical photometry of \vsco\ on the nights of 25, 26, and
27 August 2006 using the University of Cape Town CCD Photometer (UCTCCD;
\citealt{OD1995}) on the South African Astronomical Observatory (SAAO)
1.9m telescope.  The detector is a Wright
Instruments Peltier-cooled CCD with a thinned and back-illuminated EEV
P86321/T chip. It was used in frame-transfer mode where one half of the
CCD is masked and used as a readout area, allowing exposures to be made
with essentially no ``dead'' time between exposures.  On the 1.9m telescope,
the pixels on the detector correspond to 0.13 arcsec so that it is normal
to use 3x3 or greater pre-binning, unless the seeing is better than about
1 arcsec, to ensure optimal data-extraction. Because \vsco\ is relatively
bright, integration times of 6 seconds were used on all nights.  All
observations were made in white light. The CCD was pre-binned at 5x5, 4x4,
and 3x3 for the three nights respectively.

Observations were reduced at the telescope using a custom pipeline, which
allows the observer to select suitable comparison stars and judge observations.
The conventional procedure of reducing observations were followed (bias
subtraction, flatfield correction etc.) after which the stellar brightnesses
were extracted using the method described in \cite{Sea1993}.
Differential corrections were made by using the lightcurve of the brightest
nearby star. Due to the small size of the UCTCCD (50x34 arcsec$^{2}$ on the
1.9m telescope) the only available comparison stars were somewhat fainter
than \vsco\ but they were nevertheless used as the differentially corrected
lightcurves were deemed to be of better quality than the uncorrected ones. 
With this set-up, the statistical fluctuation of the
data is much less than the intrinsic variability of \vsco.  Typical number
of counts in a 6 s integration was 100,000, and it never dropped below 30,000
even during the eclipse.

\section{Results}

\subsection{X-ray Spectrum}

\begin{figure}
\plotone{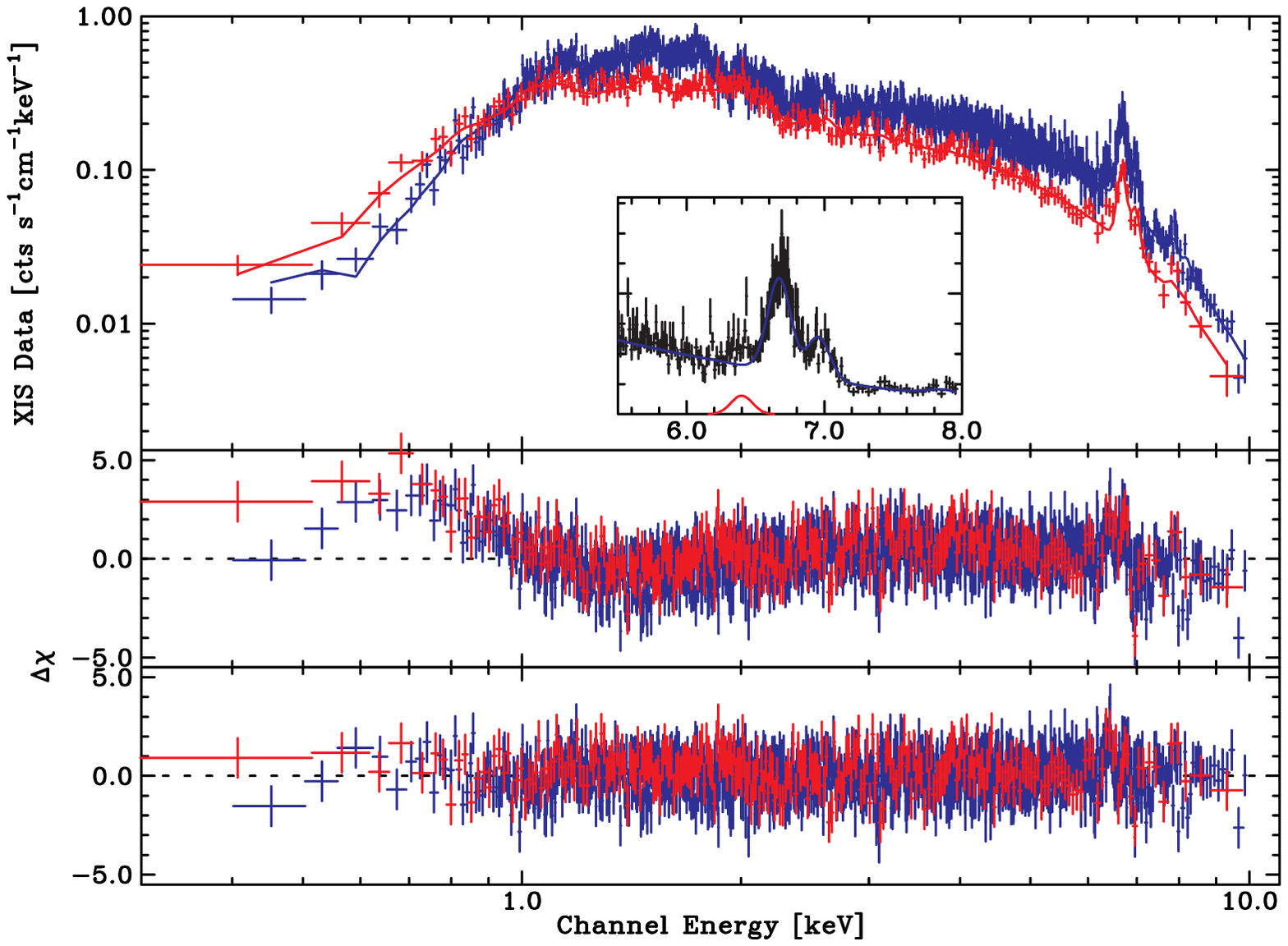}
\caption{The \suzaku\ XIS spectra (blue: sum of three units with FI
CCDs; red: XIS1 data, containing the BI CCD) of V893~Sco fitted with
the cooling flow ({\tt mkcflow}) model.  The data are shown in the top panel.
The middle panel shows the residuals (as $\sigma$s) when only a simple absorber
is used.  The bottom panel shows the residuals with a simple and a partial
covering absorber.  The solid line on the top panel shows the latter model.
The inset in the top panel gives an expanded view of the Fe K$\alpha$ complex,
here fitted with the {\tt mkcflow} model plus a 6.4 keV Gaussian.}
\label{xcont}
\end{figure}

Low resolution X-ray spectra of dwarf novae are generally dominated by
an absorbed bremsstrahlung-like continuum and a prominent He-like Fe
K$\alpha$ at 6.7 keV \citep{Bea2005,Pea2005}.  The \suzaku\ spectrum
of \vsco\ is no exception, as shown in Figure\,\ref{xcont}.  The Fe
K$\alpha$ line indicates the origin in an optically-thin, thermal plasma.
Therefore the simplest model consistent with the data is the single
temperature plasma model {\tt mekal} \citep{Mea1985,Mea1986,Lea1995,Kea1996}
as implemented in {\tt Xspec} \citep{A1996}.  There is a strong reason to
believe that this is not a physically correct model for X-ray emission
in CVs: the post-shock plasma in CVs loses energy by emitting X-rays and
therefore must cool, and cannot remain at a single temperature.
However, it is certainly possible for a multi-temperature plasma
to have a strong peak in the differential emission measure distribution
such that a single-temperature plasma model provides a good fit.
For this reason, and to allow easy comparison with other works,
we present the results of the {\tt mekal} model fit in
Table\,\ref{tab:globalfit}.

However, as our primary spectral model, we choose the {\tt mkcflow} model
that was originally developed for the cooling flow in clusters of galaxies
\citep{MS1988}, but was successfully applied to X-ray emission from CVs
\citep{Mea2003}.  In its original context, the cooling flow model assumes
a steady-state with a large reservoir of plasma in the outer parts of the
cluster at the maximum temperature, kT$_{max}$; the plasma cools by emitting
X-rays with no additional energy input.  This model is applicable to CVs
as long as the post-shock region is in a steady state and there are no
additional heating or cooling terms beyond X-ray emission.  

There are other models of multi-temperature X-ray emission specially
developed for magnetic CVs (e.g., \citealt{Cea1998,Sea2005}).  However,
as \cite{MS1988} state (section III): ``Assuming that the same mass flow
rate pertains throughout the cooling flow, the emission measure for each
temperature is determined by the time it takes for the matter to radiate
away sufficient energy to cool down to the next temperature shell.
The differential emission measure is thus proportional to the
reciprocal of the bolometric luminosity at that temperature.''
It is therefore unnecessary to solve for the structure of the emission
region in order to predict the correct output X-ray spectrum,
in the absence of additional heating or cooling terms.  Thus,
{\tt mkcflow} serves well a 1st order approximation, both in
clusters of galaxies and in CVs.

We present the results of our fit in Figure\,\ref{xcont} (for the
{\tt mkcflow} model) and in Table\,\ref{tab:globalfit} (for both models).
Note that we have used a global abundance value in our fits, and
in effect it is the Fe abundance we have measured.  This is because
the only discrete feature evident in the \suzaku\ spectrum of \vsco\ is
the Fe K$\alpha$ complex; the abundances of other elements cannot
be constrained reliably, in the absence of discrete features.
To investigate the Fe K$\alpha$ complex further, we fitted an
absorbed bremsstrahlung continuum and three Gaussians in the
5--10 keV range.  We do not detect any energy shifts or broadening
of the Fe lines, although the upper limits are unconstraining due
mainly to the modest signal-to-noise ratio of the XIS spectra.
In addition to the dominant 6.7 keV (He-like) line with an equivalent
width of 400$\pm$30 eV, we also detected both the H-like line (6.97 keV;
100$\pm$20 eV) and the fluorescent line at 6.4 keV (40$\pm$15 eV).
The small equivalent width of the 6.4 keV line indicates that optically
thick matter (white dwarf surface and an optically thick inner disk;
\citealt{DO1997}) covers less than 1$\pi$ steradian of the sky as seen
from the emission region, which would result in an equivalent width
in the 100--150 eV range \citep{GF1991}.  Although we did not include
the fluorescent line in the global fit, it is sufficiently weak not to
have affected the quality of the fit or the derived parameter values.

We note that the derived abundances depend on our choice of spectral
model (Table\,\ref{tab:globalfit}).  It is likely that a single
temperature {\tt mekal} model underestimates the Fe abundance, because
the best fit temperature is near the peak of He-like Fe K$\alpha$ emissivity.
In the {\tt mkcflow} fit, the plasma temperature range extends well
outside the peak emissivity of He-like Fe K$\alpha$, hence a higher
abundance is required to fit the observed data, and this probably
is a more realistic reflection of the true abundances of \vsco.

With both {\tt mekal} and {\tt mkcflow} models, the use of a single
absorber (using the {\tt wabs} model) resulted in significant residuals
particularly below 1 keV (see the middle panel of Figure\,\ref{xcont}).
The fitted \nh\ of $> 5 \times 10^{21}$ cm$^{-2}$ is too high to be the
interstellar column towards \vsco\ at d=155 pc; even the total Galactic
column in this direction is only 1.4$\times 10^{21}$ cm$^{-2}$ \citep{DL1990}
as reported by the HEASARC Nh
tool\footnote{http://heasarc.gsfc.nasa.gov/cgi-bin/Tools/w3nh/w3nh.pl}.
Thus, we conclude that the measured column is dominated by an intrinsic
absorber within the \vsco\ binary system.  For example, our line of sight
to the boundary layer may pass through the inner accretion disk
\citep{vTea1996}.

The magnitude of the low energy residuals is far too large to be
removed by changing the assumed abundances of the absorber.  For
example, substituting {\tt wabs} with {\tt phabs} model instead,
and trying various ``solar'' abundance values (e.g., those
according to \citealt{AG1989} and by \citealt{Wea2000}) with
the {\tt mkcflow} model resulted in essentially the same values
of reduced $\chi^2$ (1.32--1.34).  One can get an improved fit
(reduced $\chi^2$=1.09) by changing abundances element by element
using the {\tt varabs} model, but at the expense of non-astrophysical
abundances ($>$20 times solar for Na, 0 for Al through Cr, and
essentially solar for Fe).  On the contrary, the low energy residual
was successfully removed by adding an additional partial covering
absorber ({\tt pcfabs}) (see the top and the bottom panels of
Figure\,\ref{xcont} and Table\,\ref{tab:globalfit}).

We also tried using an additional emission component instead, while keeping
just a single {\tt wabs} absorber.  These were not as successful
at removing the residuals as the complex absorber: adding a low temperature
{\tt mekal} to {\tt mkcflow} resulted in $\chi^2_\nu$=1.08 with residuals
below 1 keV, and the addition of a blackbody resulted in $\chi^2_\nu$=1.04
with residuals below 0.6 keV, compared to $\chi^2_\nu$=0.766 for the complex
absorber fit.  Thus, we prefer the partial covering absorber interpretation,
also because this is a natural consequence if (as we argue below) the X-ray
emission region and the absorber are comparable in size: barring a fine-tuned
geometry, an absorber whose size is comparable to that of the emission region
cannot completely block our line of sight.

This intrinsic absorber also offers a partial explanation for the high XSS
($>$3 keV, unaffected by absorption) to RASS ($<$2 keV, strongly affected)
count ratio.  However, source variability must also play a role.  Using the
best-fit partial-covering {\tt mkcflow} model, the \suzaku\ spectrum
corresponds to a ``predicted'' \rosat\ PSPC (on-axis) count rate of 0.24 \cps.
In reality, the vignetting-corrected RASS rate is 0.35 \cps, so it was
likely brighter during the RASS epoch by a factor of $\sim$1.5.  For the
\xte\ PCA, predictions (XSS measurements) are 1.35 \cps\ (2.38) in the
3--8 keV band for a ratio of $\sim$1.7 and 0.51 \cps\ (0.96) for $\sim$1.9
in the 8--20 keV band.  This could also be due to source variability.
Another possibility is that there was an additional source within the
PCA field of view (but outside the \suzaku\ XIS field of view) during
the slew that led to the inclusion of \vsco\ in the XSS catalog.

The 2--10 keV absorbed flux of \vsco\ is 1.7$\times 10^{-11}$ \epcs,
corresponding to a 2--10 keV luminosity of 4.9$\times 10^{31}$ \eps\ at
155 pc assuming isotropic emission.  The 0.4--2 keV absorbed flux is
3.0$\times 10^{-12}$ \epcs,
while the extrapolated flux of \vsco\ above 10 keV is $< 1 \times 10^{-11}$
\epcs.  This justifies our choice not to analyze the HXD data, since a source
at this flux level is undetectable given the current systematic uncertainties
in the background reconstruction.  The estimated bolometric X-ray luminosity
depends on the choice of spectral models (see Table\,\ref{tab:globalfit}),
but may be as high as 1.4$\times 10^{32}$ \eps, placing \vsco\ among the
most X-ray luminous dwarf nova \citep{Bea2005}.

\subsection{Optical Light Curves}

We had three purposes in mind for obtaining optical photometry
of \vsco\ contemporaneously with the \suzaku\ observations.
First, we wished to ascertain the outburst status of the object.
The instrumental white light magnitudes, as measured with the SAAO
1.9 m telescope, were in the range 11.0--11.5, 11.5--12.0, and 11.2--11.8
outside the eclipse, respectively, on the nights of Aug 25, 26, and 27.
Since there is a known offset of approximately 3.0--3.2 magnitudes with
respect to the standard V band for the typical spectrum of a CV,
\vsco\ had V magnitudes of 14.0--15.0 out of eclipse on these nights.
That is, \vsco\ was in quiescence during the \suzaku\ observations.

Our second purpose was to check the orbital ephemeris.  We therefore
folded the SAAO light curves on the ephemeris of \cite{BSG2000}
and found that the eclipse occurred $\sim$0.17 cycles earlier than
predicted.  There is a narrow, V-shaped core of the optical eclipse
lasting $\sim$0.03 cycles, presumably indicating the eclipse of a compact
light source such as the white dwarf photosphere.  We therefore measured
the times of the fourth contact, as indicated by the change of slope of the
light curve, some 0.015 cycle after the mid-eclipse.  We also measured
the times of the first contact, indicated by the slight change in slope
during ingress, $\sim$0.015 cycle before the mid-eclipse, even though
there often was ambiguity exactly where this happened.  The results for
the 6 eclipses (2 were observed on each of the three nights) are listed
in Table\,\ref{tab:contacts}, along with the brightness level relative
to the faintest point during that eclipse.  On average, we measured
a slight steepening of the gradient at Bruch et al. phase 0.8093$\pm$0.005,
$\sim$0.56 mag above mid eclipse; and the end of the steep part of the
egress at Bruch et al. phase 0.8385$\pm$0.0011, $\sim$0.47 mag above.
We therefore applied a offset of $-$0.176 to the phases calculated
using the \cite{BSG2000} ephemeris.  This is equivalent to adopting a
linear ephemeris with a period of 0.075961467 d (cf. 0.07596185$\pm$0.00000012
d of the ephemeris) while keeping the same epoch for cycle 0, although
the time gap of 7 years between the Bruch et al. observation and ours
is long enough that a quadratic ephemeris may be more realistic.

The final purpose of the optical photometry was a detailed comparison
of the X-ray and optical light curves.  We will present our analysis
of the X-ray light curves in \S 3.3, and report on our
modeling of both the optical and X-ray light curves in \S 4.3.

\subsection{X-ray Light Curves}

\begin{figure}
\plotone{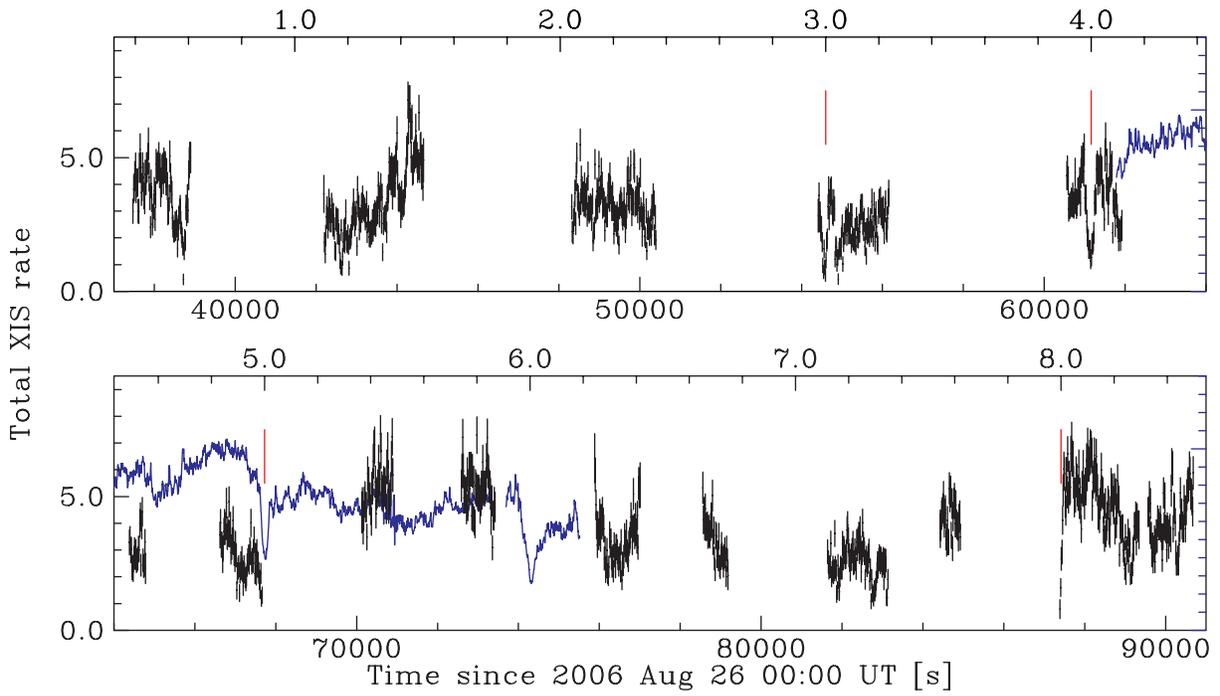}
\caption{The X-ray light curve (black) of \vsco\ in 16 s bins plotted
against time (bottom axis) and, equivalently, relative cycle counts (top
axis).  The Aug 26 SAAO photometry is shown in blue.  Red tick marks shows the
two full and two partial coverage of the eclipse phase by the \suzaku\ XIS.}
\label{xlc}
\end{figure}

We present the \suzaku\ light curve of \vsco\ against both time and relative
cycle counts (since cycle 34890.0 of Bruch et al. ephemeris, adjusted as
described in \S 3.2) in Figure\,\ref{xlc}.  Due to the restrictions of
low Earth orbit satellites, the \suzaku\ coverage has many gaps, resulting
unfortunately in little strict overlap with the SAAO photometry.  The phase
coverage is highly uneven: it can be seen that phase interval 0.6--0.9,
for example, is poorly covered.  The \suzaku\ count rate is variable at
all orbital phases, but there are dips whenever the phase of the optical
eclipse is covered.  The \suzaku\ data cover the eclipse phase completely
twice (at relative cycle count 3.0 and 4.0); in addition, they cover a part
of the ingress once (5.0, as the satellite was approaching the night-side
limb of the Earth), and the egress once (8.0, as the satellite emerged from
the daylight limb of the Earth).

\begin{figure}
\plotone{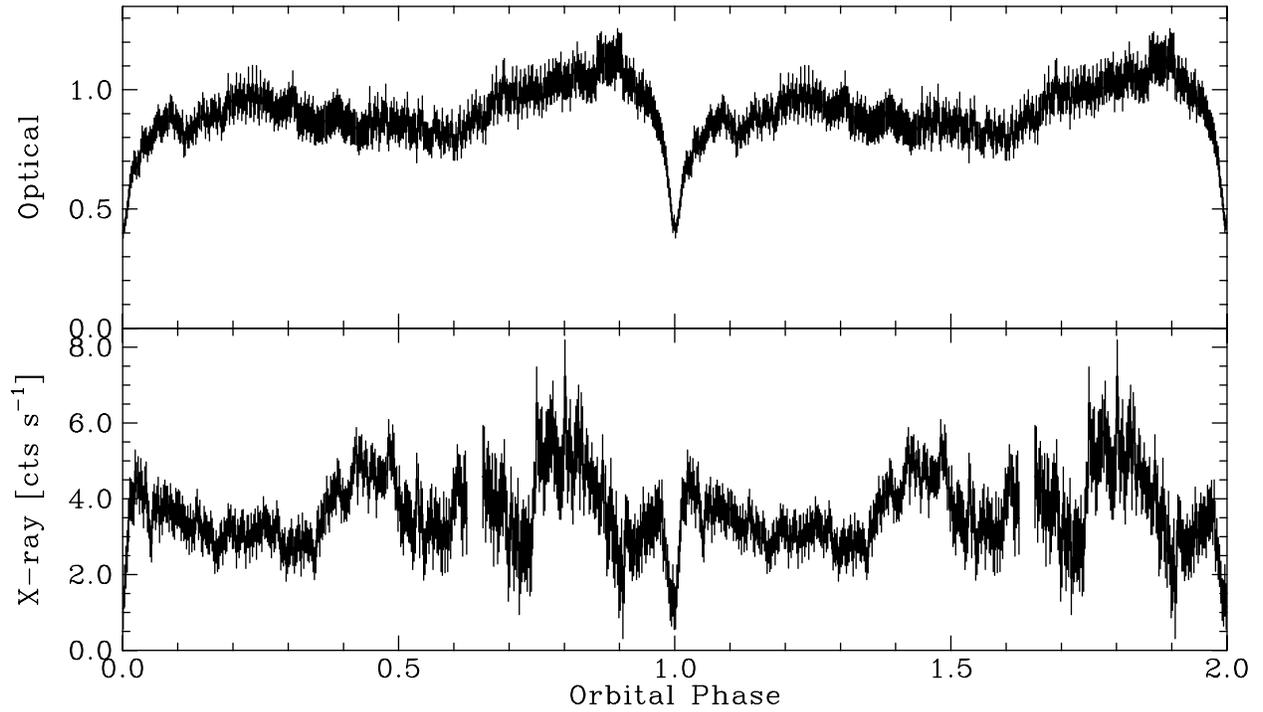}
\caption{Folded light curves of V893~Sco from our SAAO (top) and
\suzaku\ (bottom) observations.  Two complete cycles are shown for
clarity.  Optical data have been converted to linear intensity scale,
with a normalization such that 1.0 roughly corresponds to V=14.5,
or $\sim$125,000 counts in a 6 s integration.}
\label{ecl}
\end{figure}

In the folded representation (Figure\,\ref{ecl}), the X-ray light curve
shows a significant dip at the phase of the optical eclipse.  It is
also clearly variable at other phases, but no other dips of comparable
significance are seen.  Given the poor phase coverage of the \suzaku\ data,
it is premature to consider any orbital
modulations outside the eclipse phase.  The apparently greater amplitude
of variability during the second half of the orbital cycle (0.5--1.0), in
particular, is probably due to the poorer coverage of these phases.
Concentrating on the dip at phase 0.0, we investigated its properties
in more details to ascertain if it is an eclipse.

\begin{figure}
\plotone{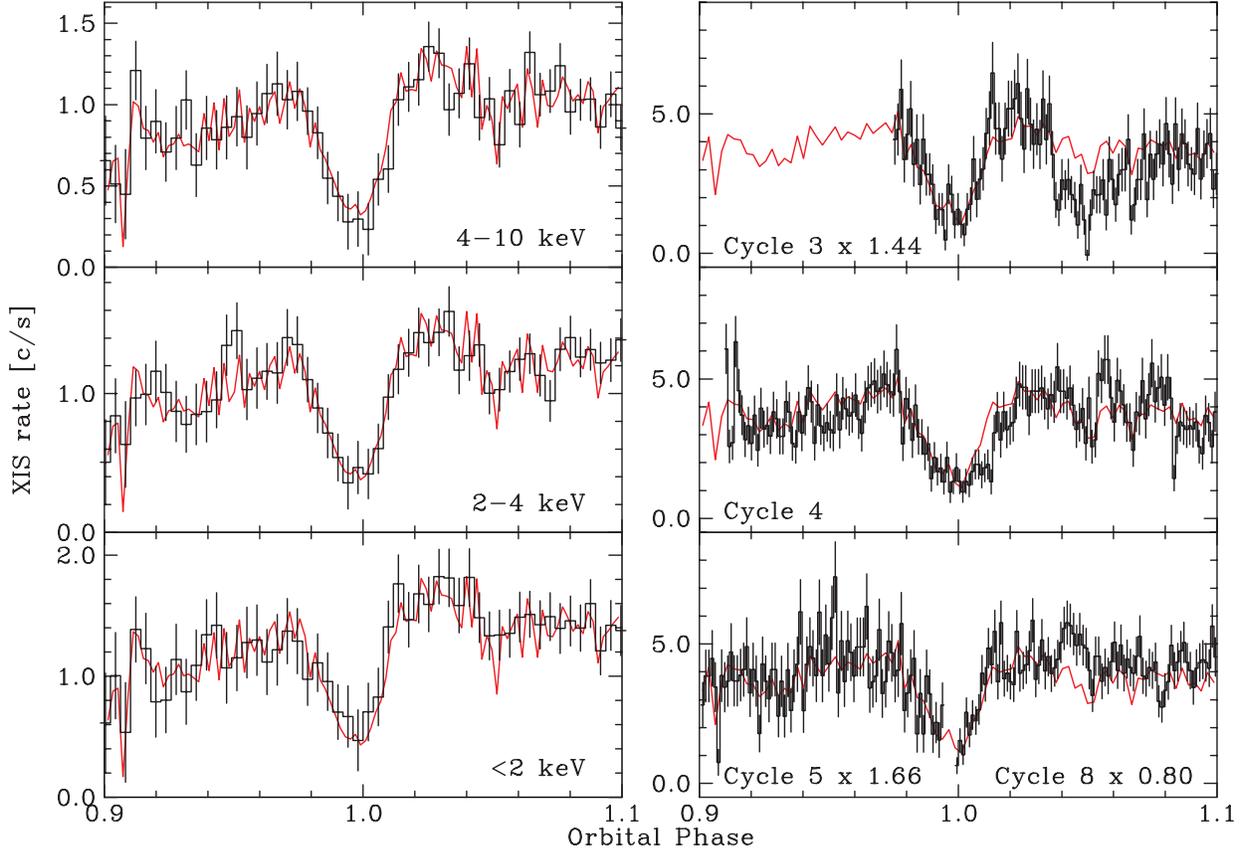}
\caption{(Left) The folded X-ray light curves near the eclipse phase
in three energy ranges.  In each panel, the black histograms show the data
in the energy range shown, while the red line is the total (0.3--10 keV)
light curve.  All curves are scaled such that the top of the panel is 1.2
times the largest bin. (Right) The individual X-ray light curves around
the eclipse phase.  In each panel, the black histograms show an individual
cycle (cycles 5 and 8 are shown together in the bottom panel).  Those of
cycles 3, 5 and 8 are scaled by the factors shown and explained in the text.
Red solid line, repeated in all 3 panels, is the average of these normalized
light curves.}
\label{twobythree}
\end{figure}

First, we investigated if this dip is energy dependent.  In the left half
of Figure \,\ref{twobythree}, we show the folded light curve near
the eclipse phase in three energy bands, as histograms.  Solid lines,
repeated in all three panels, are the eclipse profile for the total band.
Both the individual and the total-band light curves are simply scaled
so that the top of the panel corresponds to 1.2 times the maximum
value of all phase bins for that curve.  It is clear from these panels
that there is little, if any, energy dependence in the depth of this dip.
We therefore conclude that this is not an absorption event, e.g., by the
outer rim of the disk, or by an extended atmosphere of the secondary.

We also investigated how repeatable the dip was from cycle to cycle.
A comparison between the average profile (shown in Figure\,\ref{ecl})
and individual cycles would be dominated by the slow ($>$1,000 s)
variations in the source brightness.  For each segment around the
eclipse phase, we first calculated the average intensities during
phases 0.96--0.98 and 0.02--0.04.  We then normalized the individual
light curves using the ratio of these ``out-of-eclipse'' averages,
relative to cycle 4.  We also calculated the average of the light
curves thus normalized.  In the right half of Figure \,\ref{twobythree},
we show both the normalized individual profiles (black histogram with
error bars) and their average (red line).  We confirm that
the reduction in count rate is seen every time the eclipse phase is observed.
At the same time, there are variations in the light curves from cycle to
cycle.  This is most apparent in the egress for cycle 4, for which two
interpretations are possible.  Either the egress was delayed by $\sim$0.008
cycle or the total emission was lower by 30--40\% during the egress.
Given the small number of cycles observed, we must keep in mind
the possibility that the ingress is similarly affected by a similar
cycle-to-cycle variability.

To summarize, we observe a significant decline in the X-ray count rates
during the phase of the optical eclipse (Figure\,\ref{ecl}).  This
feature is present every cycle, even though there are variations in
the detailed profile from cycle to cycle, and the fractional decrease
is independent of the photon energy (Figure\,\ref{twobythree}).  We
conclude that we have discovered a partial eclipse of the X-ray emitting
region by the secondary in \vsco.

\section{Discussion}

\subsection{Intrinsic Absorber, Abundances, and Luminosity}

The X-ray spectrum of \vsco\ indicates the presence of a partial-covering
intrinsic absorber (Table\,\ref{tab:globalfit}) with \nh\ 
$\sim 2 \times 10^{22}$ cm$^{-2}$.  Such an absorber reduces the observed
count rate in the \rosat\ band, but has little effect above 2 keV.
Our finding is consistent with the study of \cite{vTea1996}, who find
that intrinsic absorption in the \rosat\ band is a general phenomenon
among high inclination non-magnetic CVs.  They conclude that this absorber
is likely located in the inner disk near the boundary layer.  Since the
innermost region of the accretion disk has a similar physical dimension
as the boundary layer, partial-covering absorption is a likely result.

This interpretation is in conflict with the predictions of the
basic version of the disk instability model (see, e.g., \citealt{L2001})
that typically predicts a surface density of the inner disk of order
1 g\,cm$^{-2}$.  Viewing the putative boundary layer through such an
inner disk should results in \nh\ $\sim 10^{24}$ cm$^{-2}$.  If this
is true, the inner disk is optically thick to $>$7 keV photons,
and will produce reflection signatures, including a strong 6.4 keV
line \citep{DO1997}.  The weakness of the observed 6.4 keV line in
\vsco\ excludes the possibility that such a dense ($\sim$1 g\,cm$^{-2}$)
disk subtends a large ($\sim 2\pi$ steradian) solid angle.  Our discovery
of a $\sim 2 \times 10^{22}$ cm$^{-2}$ partial covering absorber in
\vsco\ suggests that the disk may be far less dense than the basic
version of the disk instability theory predicts.  Note that
this same model is already in conflict with other key observations,
such as the interoutburst intervals and the quiescent X-ray luminosity
\citep{L2001}.

We note, in passing, that \cite{Rea2008} classified three CVs as
likely IPs based largely on the complex absorber found in their
X-ray spectra.  Based on our result on \vsco, as well as the similar
finding on OY~Car \citep{Pea2005}, we urge caution against using X-ray
absorption alone to claim an IP classification.

We infer that the bolometric X-ray luminosity of \vsco\ to be
1.0--1.4$\times 10^{32}$ \eps\ at the time of the \suzaku\ observations.
\vsco\ therefore appears to be among the most X-ray luminous dwarf novae.
Byckling et al. (in preparation) discuss the X-ray luminosity function
and temperatures of nearby dwarf novae, including \vsco.

\subsection{The Partial X-ray Eclipse}

The detection of a partial eclipse in a non-magnetic CV is unprecedented.
All previous observations of deeply eclipsing, quiescent dwarf novae
showed flat-bottomed X-ray eclipses that are consistent with being total,
perhaps except for a weak residual contribution from the corona of the
secondary \citep{Wea1995,vT1997,Mea1997,Pea1999,Rea2001,WW2003}.
The transitions into and out of eclipses have been used to constrain the
total size of the X-ray emission region to be not much larger than the
white dwarf.  On the other hand, the partial X-ray eclipse of \vsco\ is
also in marked contrast to those of U~Gem \citep{Sea1996} and
WZ~Sge \citep{Pea1998}.  These are both lower inclination systems that
exhibit a partial eclipse of the outer accretion disk and the bright
spot in the optical.  There is no X-ray eclipse in these systems, instead
showing energy-dependent absorption dips prior to the eclipse.
These dips are believed to be caused by photoelectric absorption in
an azimuthal structures on the outer accretion disk.

X-ray emission from regions other than the vicinity of the white
dwarf and from the corona of the secondary have been considered
in the literature. One speculated source is the bright spot region. 
However, this possibility was comprehensively refuted by \cite{P1977};
in particular, the highest shock temperature possible for the bright spot
is of order kT$\sim$1 keV, and therefore this can be discarded for \vsco. 
The other source is the soft X-ray component observed in high accretion
rate systems, whose physical origin is unclear.  For example, in the
\xmm\ observations of the eclipsing nova-like system, UX~UMa, \cite{Pea2004}
found two components of X-ray emission. One is highly absorbed
(\nh\ $\sim 6 \times 10^{22}$ cm$^{-2}$) thermal (a single temperature
{\tt mekal} fit gives kT=5.5 keV) component and undergoes a total
eclipse, and therefore originates from a compact region around the white
dwarf.  The other component that dominates below 2 keV unabsorbed and
does not show total or partial eclipse, and is therefore from a highly
extended region. However, this component has the wrong spectral shape
and its origin is far too extended to offer an explanation for the
partially eclipsed component in \vsco.  We therefore seek an explanation
of the partial X-ray eclipse assuming a compact X-ray emitting region
centered on the white dwarf.

We believe that a grazing geometry in which the entire photosphere of the
white dwarf is never completely eclipsed can explain both the optical
and the X-ray eclipse profiles.  To demonstrate that such a geometry is not
too contrived, let us consider a plausible set of system parameters and
estimate the probability of a partial eclipse of the white dwarf.
Given the 1.82 hr orbital period, an assumed mass ratio of 0.25,
an empirical mass-radius relationship of the secondary \citep{P1984},
and a white dwarf mass-radius relationship \citep{PW1975}, we estimate
a secondary radius R$_2$ of 1.28$\times 10^{10}$ cm, a primary radius
R$_1$ of 8.86$\times 10^8$ cm, and a binary separation $a$ of
4.73$\times 10^{10}$ cm (see Table\,\ref{tab:system}).  With these
representative numbers, and approximating the Roche-lobe filling
secondary as a sphere, the bottom of the white dwarf is just eclipsed
at an inclination angle $i$ of 73.88$^\circ$; the top at $i$=75.88$^\circ$,
compared to the range of 75.88$^\circ$--90$^\circ$ for a total eclipse.
The key factor is the ratio of the white dwarf diameter to the radius of the
secondary, which is $\sim$1/7 in this case, and this should be the
approximate ratio of systems with partial and total eclipses of the
white dwarf photosphere.

All available data are consistent with our interpretation that
the white dwarf in \vsco\ is not totally eclipsed.  A total eclipse
would result in a flat-bottomed optical eclipse.  Systems with flat-bottomed
light curves are often indicated in catalogs as having a double eclipse:
this is the terminology for eclipses that show two episodes of rapid ingress
and egress, one caused by the total eclipse of the white dwarf and the
second by that of the bright spot, offset in phase because the bright
spot is located away from the line of centers of the binary.  The white
dwarfs in double-eclipsing systems must undergo a total eclipse, because
otherwise it would be impossible to separate the ingress of the white
dwarf from that of the bright spot.  On the other hand, the optical
eclipse in \vsco\ is neither flat-bottomed or double (\citealt{BSG2000};
this work).  Therefore, if the white dwarf is eclipsed at all,
it must be a partial eclipse.

\begin{figure}
\plotone{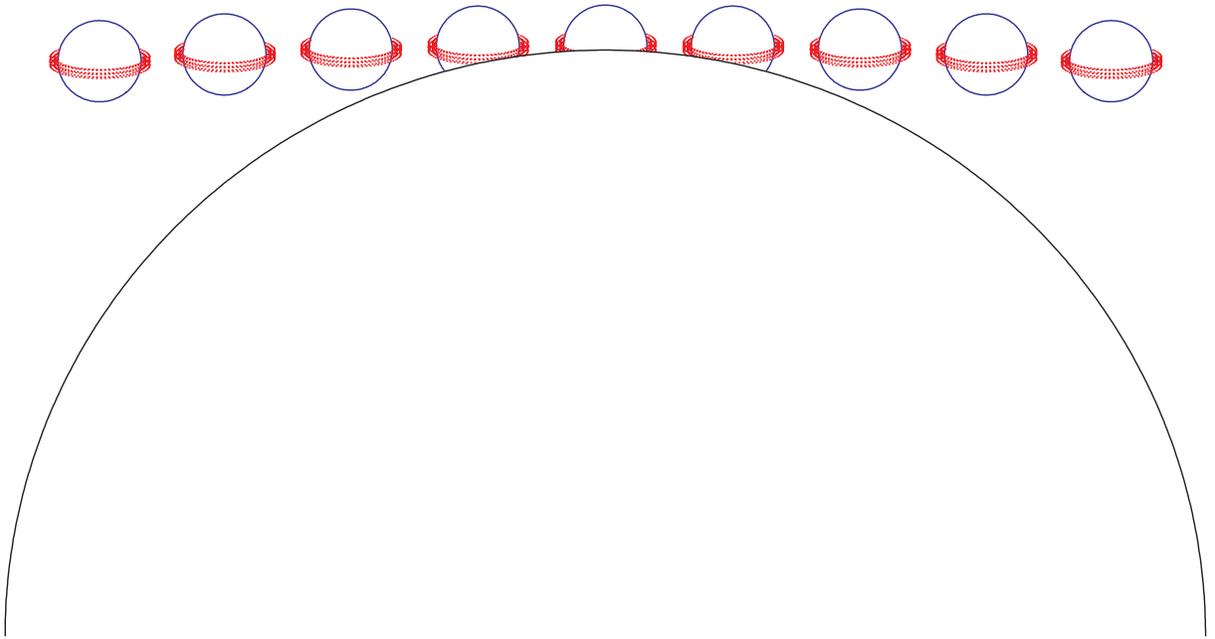}
\caption{A schematic diagram of the possible geometry of the grazing
eclipse in V893 Sco.}
\label{schema}
\end{figure}

We now consider the X-ray eclipse in such a system assuming three different
geometries of the X-ray emission region:  Magnetic pole region(s) as
appropriate for IPs, a spherical corona that covers the entire white
dwarf photosphere, and an equatorial boundary layer.  We sketch a possible
geometry for the boundary layer case in Figure\,\ref{schema}.

The accretion in IPs, by definition, occurs
onto an area smaller than the entire surface of the white dwarf.  For
the IP, XY~Ari, \cite{H1997} used the rapidness of the eclipse ingress
($<$2 s) and their phase jitters (individual orbital phases of the third
contact varied in the range 0.0543--0.0553) to infer small ($<$0.002 of
the total surface area) spots that move on the projected face of the white
dwarf as a function of the spin phase.  It's extremely unlikely that such
a small spot be partially eclipsed by the secondary.  If, on the other
hand, two larger spots are located near the rotational poles, a
flat-bottomed partial eclipse is a likely result, as we do indeed observe
in EX~Hya \citep{Rea1991}. Based on the observed X-ray eclipse profile,
and the lack of phase jitters, we believe that an IP or IP-like geometry
is highly unlikely for \vsco.

Next, we consider a spherical corona surrounding the white dwarf.
This simple scenario has the advantage that, if the white dwarf
photosphere is partially eclipsed, so must be the corona.  Because
such a corona is bigger then the white dwarf in every direction,
the X-ray eclipse must last longer.  The depth of the X-ray eclipse
should be similar to that of the white dwarf photosphere, if the
latter can be reliably extracted from the optical eclipse.  Finally,
the X-ray eclipse should be U-shaped, from the limb brightening of an
optically thin emission plus the fact that the body of the white dwarf
blocks the emission from the far side.  Although none of these predicted
indicators is seen in \vsco, the quality of the current data quality
cannot rule out a spherical corona model.

Finally, we consider the boundary layer picture.  We must first consider
whether a partial eclipse of an equatorial boundary layer is likely.
This turns out not to be a severe problem.  The reason for this is that,
due to the inclination of $\sim 75^\circ$, the projected equator of the
white dwarf is significantly curved, spanning $\sim$1/8 of the white dwarf
diameter.  Thus, even a smallest possible boundary layer of a
negligible height and a negligible extent in latitude will suffer a
partial eclipse in $\sim$1/8 of the partial eclipsing dwarf nova.
The probability can only rise with any spatial extent of the boundary layer.

\begin{figure}
\plotone{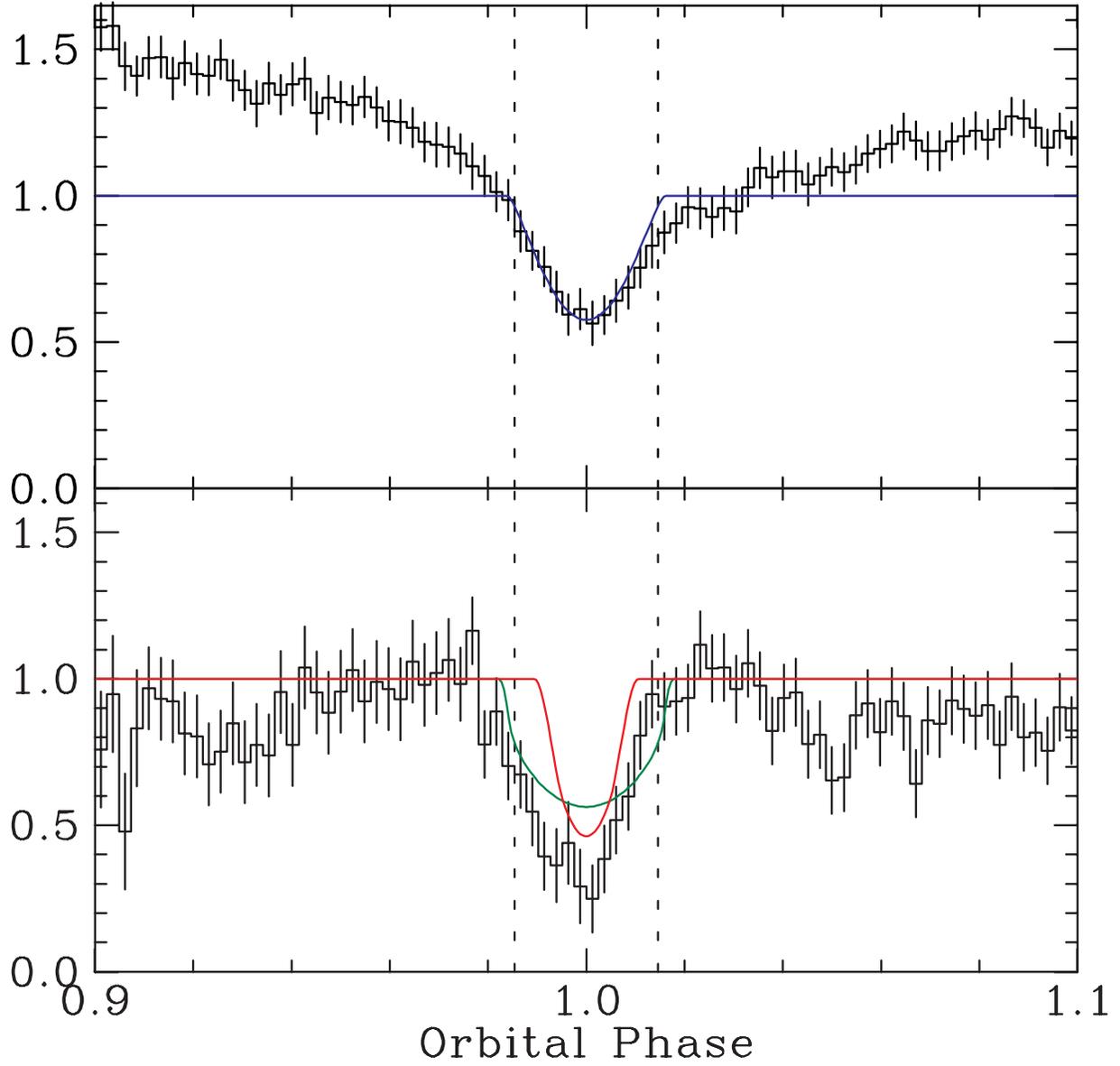}
\caption{Optical (top) and X-ray (bottom) eclipse light curves plotted with
         representative models.  The folded optical light curves are normalized
	 at the estimated time of the first contact of the white dwarf.  The
	 X-ray light curve is normalized so that the average rate at phases
	 0.96--0.98 and 0.02--0.04 is 1.0.  For the details of the model
         parameters, see text.}
\label{profile}
\end{figure}

As a pilot study to demonstrate that a partial eclipse interpretation
is valid and that future investigations hold the potential to constrain
the shape and the size of the X-ray emission region, we have created a
numerical model of the optical and X-ray eclipse.  We represent the
white dwarf photosphere, a boundary layer, or a spherical corona using
a set of grid points.  We trace the line-of-sight from each point at
various orbital phases to determine whether it is visible from Earth, blocked
by the body of the white dwarf, or eclipsed by the secondary.  We use the
same code to produce the schematic diagram (Figure\,\ref{schema}) and the
model light curves (Figure\,\ref{profile}).

For the optical eclipse, we plot the folded optical light curve,
normalized to 1.0 at the phase of the first contact of the compact
light source (at Bruch et al. phase 0.8093; Table\,\ref{tab:contacts}).
In this initial attempt, we only model the eclipse of the white dwarf
photosphere, since there is little information on the location of the
bright spot to allow us to construct a reliable model.  The eclipse light
curve of the photosphere is estimated at 2,000 phase bins.  The Earth
facing side of the photosphere is divided into 90 zones from the center
to the limb, and 360 points in each zone.  The parameters of the light
curve model are the R$_1$/$a$, R$_2$/$a$, and the inclination angle $i$.
These parameters can be narrowed down using astrophysical considerations.
For the secondary, we have assumed the empirical mass-radius relationship
of \cite{P1984} as before (Table\,\ref{tab:system}).
We then assumed a mass ratio $q$ = M$_2$/M$_1$ = 0.25, which determines
the white dwarf mass, and hence the radius.  We then stepped through a
range of $i$ to reproduce the phases of the first and fourth contacts
(Table\,\ref{tab:contacts}).  We find an inclination of $i$=74.2$^\circ$
under these assumptions.  The model shown reflects the additional
assumption that the white dwarf was the dominant light source at first
and forth contacts, and that other sources of light are either too faint
(likely to be the case for the secondary and possibly also for the
accretion disk) or already fully eclipsed.  Different conclusions would
result if we assume a different value of $q$, or if the narrow eclipse
is a combination of white dwarf and bright spot eclipse.  As for limb
darkening, we have also simulated a case with an extreme limb darkening
($u$=1 following, and using the notation of, \citealt{WH1990}), but this
does not make a large enough difference to matter here.  The vertical
dashed lines in Figure\,\ref{profile} are the ingress and egress timings
we measured and used to adjust the orbital ephemeris (\S3.2 and
Table\,\ref{tab:contacts}).

For the X-ray light curve, we use the same normalized and averaged version
as shown in Figure\,\ref{twobythree}.  For the spherical corona model
(shown in green), the parameters are R$_1$/$a$, R$_2$/$a$, $i$, and
the height of the corona above the white dwarf $h$.  The corona is
divided into 10 layers in height, 180 degrees of latitude, and 360
degrees of longitude, and assumed to emit uniformly throughout this
volume.  The system parameter values are the same as those used for
the optical eclipse, and we assume a height of 0.2 R$_1$.  Changing
the height will only affect the width of the X-ray eclipse under the
assumption of a spherical corona.

For the boundary layer, we assumed a boundary layer that extends
$\pm 5^\circ$ from the equator in latitude (divided into 21 zones),
and an X-ray emission region height of 0.2R$_1$ (10 layers), which
resulted in the red curve.  Both the height and the latitudinal extent
affects the depth and the width of the predicted eclipse, but
in different directions in the width-depth plane.

These preliminary models reproduce the duration and the depth of
both the optical and X-ray eclipses to better than a factor of 2.
To proceed to the next level of actually fitting the optical and X-ray
light curves, several things must happen.  First, the system parameters
must be constrained from UV, optical, and IR observations to narrow down
the parameter space.  The mass ratio $q$ is a crucial parameter: once we
know $q$, the eclipse light curves will in principle allow us to tightly
constrain $i$.  It also provides a tight constraint on the white dwarf mass,
and hence its radius.  Knowing $q$ also fixes the initial trajectory
of matter leaving the L$_1$ point, hence constrains the possible
location of the bright spot.  Alternatively, UV spectroscopy has
the potential to constrain the white dwarf parameters \citep{Gea2008}.
UV photometry may better isolate the white dwarf eclipse from the bright
spot eclipse than optical photometry, and allow us to confirm or refute
the partial eclipse of its photosphere.

Second, the quality of the X-ray light curve must improve.  The statistical
quality is one issue, but we also need to observe an increased number of X-ray
eclipses to average out cycle-to-cycle variations.  Observations from
high Earth orbit will help in allowing a better modeling of the longer-term
(minutes to hours) variability and to separate that from the
eclipse profile.

\section{Conclusions}

We performed \suzaku\ observations of the eclipsing dwarf nova,
\vsco.  We confirm that it is an X-ray luminous dwarf nova, and
moreover, report the discovery of a partial X-ray eclipse.

In the past, both optical and X-ray observers have concentrated on systems
that exhibit a total eclipse of the white dwarf.  They provide a strong
constraint on the height of the X-ray emission region above the white
dwarf. However, eclipse light curves are one-dimensional, and one can
only derive one-dimensional constraints from the eclipse analysis.
The X-ray emission region, on the other hand, has two dimensions
(height and latitudinal extent) assuming azimuthal symmetry.
It is therefore necessary to observe an ensemble of eclipsing dwarf
novae, in which the limb of the secondary cuts across the
emission region from different angles, to be able to constrain the
height and the latitudinal extent simultaneously.  Inclusion of
a partially eclipsing system in such a study will be a huge step
forward in our quest to constrain the X-ray emission region size
in two dimensions.  Future X-ray observations of \vsco, together
with improved estimates of system parameters from other wavelengths,
have the potential to test the boundary layer picture of X-ray emission
in quiescent dwarf novae, and furthermore to constrain the latitudinal
extent of such a boundary layer.

\acknowledgments

This research has made use of data obtained from the \suzaku\ satellite,
a collaborative mission between the space agencies of Japan (JAXA) and
the USA (NASA).

\clearpage

% Figure captions
%---------------------------------------------------------

\clearpage

\begin{deluxetable}{lrrrrrrrl}
\rotate
%\footnotesize
\tablecaption{Spectral Parameters of \vsco.\label{tab:globalfit}}
%\tablewidth{0pt}
\tablehead{\colhead{Model} & \colhead{$\chi^2_\nu$} &
           \colhead{N$_{\rm H}$\tablenotemark{a}} &
           \colhead{N$_{\rm H,2}$\tablenotemark{b}} &
           \colhead{CF\tablenotemark{b}} &
           \colhead{kT/kT$_{max}$} & \colhead{Abund.\tablenotemark{c}} &
	   \colhead{L$_x$\tablenotemark{d}} \\
	    & & \colhead{(10$^{21}$ cm$^{-2}$)} &
           \colhead{(10$^{22}$ cm$^{-2}$)} & & (keV) & &
           \colhead{(ergs s$^{-1}$)} }

\startdata
mekal & 1.349 & 5.20$^{+0.16}_{-0.15}$ &
      & & 9.34$^{+0.33}_{-0.32}$ & 0.87$^{+0.05}_{-0.09}$ &
%      3.04$\times 10^{-12}$ & 1.71$\times 10^{-11}$ &
      1.07$\times 10^{32}$ \\
mekal & 0.951 & 2.86$^{+0.23}_{-0.22}$ &
      1.96$^{+0.24}_{-0.21}$ & 0.51$^{+0.03}_{-0.02}$ &
      6.85$^{+0.27}_{-0.26}$ & 0.61$^{+0.05}_{-0.04}$ &
%      3.06$\times 10^{-12}$ & 1.73$\times 10^{-11}$ &
      1.15$\times 10^{32}$ \\
mkcflow & 1.349 & 6.46$^{+0.12}_{-0.12}$ &
      & & 32.4$^{+2.9}_{-1.4}$ & 1.08$^{+0.08}_{-0.07}$ &
%      1.65$^{+0.12}_{-0.16} \times 10^{-11}$ &
%      2.97$\times 10^{-12}$ & 1.75$\times 10^{-11}$ &
       1.42$\times 10^{32}$ \\
mkcflow & 0.928 & 2.68$^{+0.29}_{-0.24}$ &
      1.85$^{+0.24}_{-0.15}$ & 0.60$^{+0.04}_{-0.04}$ &
      14.5$^{+1.1}_{-1.6}$ & 0.73$^{+0.04}_{-0.06}$ &
%      3.28$^{+0.30}_{-0.26} \times 10^{-11}$ &
%      3.04$\times 10^{-12}$ & 1.74$\times 10^{-11}$ &
      1.28$\times 10^{32}$ \\

\enddata

\tablenotetext{a}{The equivalent hydrogen column density of the simple absorber}
\tablenotetext{b}{The column density and the covering fraction of the partial
                  covering absorber}
\tablenotetext{c}{The global abundance of the plasma, assuming the Solar
                  values of \cite{AG1989}}
\tablenotetext{d}{Extrapolated X-ray luminosity in the 0.01--100 keV range,
                  which is a good approximation of the bolometric X-ray
		  luminosity, assuming a distance of 155 pc.}

\end{deluxetable}

\begin{deluxetable}{lllll}
%\rotate
%\footnotesize
\tablecaption{Optical Eclipse Timings of \vsco.\label{tab:contacts}}
%\tablewidth{0pt}
\tablehead{\colhead{Eclipse} & \colhead{Transition} &
           \colhead{HJD} & \colhead{Decline (mag)} &
           \colhead{Bruch et al. Phase} }

\startdata
Aug 25-1 & Ingress & 2453973.29557 & 0.63 & 0.81096 \\
Aug 25-1 & Egress  & 2453973.29786 & 0.51 & 0.84113 \\
Aug 25-2 & Ingress & 2453973.37131 & 0.54 & 0.80805 \\
Aug 25-2 & Egress  & 2453973.37367 & 0.41 & 0.83913 \\
Aug 26-1 & Ingress & 2453974.28288 & 0.55 & 0.80841 \\
Aug 26-1 & Egress  & 2453974.28531 & 0.64 & 0.84041 \\
Aug 26-2 & Ingress & 2453974.35892 & 0.56 & 0.80946 \\
Aug 26-2 & Egress  & 2453974.36122 & 0.45 & 0.83963 \\
Aug 27-1 & Ingress & 2453975.27046 & 0.57 & 0.80936 \\
Aug 27-1 & Egress  & 2453975.27243 & 0.48 & 0.83528 \\
Aug 27-2 & Ingress & 2453975.34644 & 0.40 & 0.80965 \\
Aug 27-2 & Egress  & 2453975.34839 & 0.48 & 0.83525 \\
             &         &               &        &        \\
Average      & Ingress &               & 0.56$\pm$0.02  & 0.8093$\pm$0.0005 \\
Average      & Egress  &               & 0.47$\pm$0.04  & 0.8385$\pm$0.0011 \\
\enddata
\end{deluxetable}

\begin{deluxetable}{ccccccc}
%\rotate
%\footnotesize
\tablecaption{Assumed System Parameters\label{tab:system}}
%\tablewidth{0pt}
\tablehead{\colhead{Mass Ratio\tablenotemark{a}} &
           \colhead{Inclination\tablenotemark{a}} &
           \colhead{Separation\tablenotemark{b}} &
           \colhead{M$_1$\tablenotemark{c}} &
           \colhead{R$_1$\tablenotemark{c}} &
           \colhead{M$_2$\tablenotemark{d}} &
           \colhead{R$_2$\tablenotemark{d}} }

\startdata
0.25 & 74.2 & 4.73$\times 10^{10}$ cm & 0.58 M$_\odot$ &
8.86$\times 10^8$ cm & 0.15 M$_\odot$ & 1.28$\times 10^{10}$ cm \\
\enddata

\tablenotetext{a}{Mass ratio is assumed.  Inclination is then derived by
                  assuming that the white dwarf eclipse has a duration
		  from the first contact to the fourth of 0.0292 cycle.}
\tablenotetext{b}{Separation is calculated using the orbital period and the
                  masses of the two components.}
\tablenotetext{c}{White dwarf mass is derived from the secondary mass and
                  the assumed mass ratio; its radius is then derived using
		  the mass-radius relationship.}
\tablenotetext{d}{Secondary mass and radii are derived from the orbital
                  period assuming the empirical mass-radius relationship
		  of \cite{P1984}.}

\end{deluxetable}

\end{document}